# Line-scanning Brillouin microscopy with multiplexed two-stage VIPA spectrometer

CHENJUN SHI AND JITAO ZHANG*

*Department of Biomedical Engineering, Institute for Quantitative Health Science & Engineering, Michigan State University, East Lansing, MI 48824, USA*
*zhan2399@msu.edu*

**Abstract:** Confocal Brillouin microscopy enables high-resolution mechanical imaging but has low acquisition speed, partly due to its pixel-by-pixel mapping strategy. Line-scanning Brillouin microscopy (LSBM) significantly improves imaging speed by utilizing a multiplexing approach. However, current method is limited to a single-stage virtually imaged phased array (VIPA) spectrometer with insufficient capability of suppressing noise. Consequently, an absorptive gas chamber is often used to help reject excessive elastically scattered light. This approach requires specific tunable laser sources whose frequencies (e.g., around 780 nm) are locked to the absorption line of the gas chamber. Here, we developed a multiplexed Brillouin spectrometer for LSBM that increased the noise suppression to 57 dB without using any gas chamber. This is achieved by cascading two VIPA etalons with parallel dispersion axes in the spectrometer, where the first VIPA acts as a band-pass filter and the second as spectrum analyzer. We demonstrated its performance by acquiring Brillouin images of bio-printed phantoms with an inverted co-axial LSBM. This gas-chamber-free approach can expand the implementation of LSBM to other wavelengths where Brillouin scattering is more efficient and commercial laser sources are readily available.

## 1. Introduction

Spontaneous Brillouin light scattering refers to the interaction between photons and the intrinsic phonons, which arise primarily from thermal fluctuations within a material. This interaction induces a frequency shift in the scattered light, known as the Brillouin shift, which carries the viscoelastic information of the material [1]. Building on this phenomenon, confocal Brillouin microscopy (CBM) has emerged as an established modality for non-contact and non-invasive mechanical imaging [2-4]. In the past decade, CBM has been validated across a wide range of biological studies, ranging from cellular biomechanics [5-10] to tissue-level biomechanics [11-17]. Nevertheless, the typical acquisition speed of CBM ranges from 20–200 ms per pixel, resulting in total imaging time of about 5 minutes for a single 2D slice of a cell and more than half an hour for larger samples. This low acquisition speed not only compromises temporal resolution but also increases the risk of laser-induced phototoxicity and the potential perturbation of biological activities.

The limited imaging speed of the CBM is due to both the small number of scattered photons in the spontaneous process and the pixel-by-pixel mapping strategy. To overcome this limitation, stimulated Brillouin microscopy has recently been developed to significantly boost the number of Brillouin photons. Benefiting from the much higher scattering efficiency in the stimulated process, stimulated Brillouin microscopy achieves acquisition speeds of 0.2–20 ms per pixel [18-21]. On the other hand, within the spontaneous regime, the imaging speed can be improved through optimized mapping strategies. For example, full-field Brillouin microscopy has been developed to acquire a 2D Brillouin image in a single shot, with an equivalent acquisition time of 0.1–1 ms per pixel [22-24]. In addition, line-scanning Brillouin microscopy (LSBM) enables multiplexed acquisition of Brillouin spectra from many pixels on an illumination line, improving the equivalent imaging speed to ~1 ms per pixel [25-27]. Compared with the other approaches, LSBM requires only minor modifications to a standard CBM setup, offering a much simpler optical configuration.

In biomedical experiments, the Brillouin signal is often overwhelmed by non-Brillouin background noise, which primarily originates from the elastic scattering and interface reflections. Therefore, high noise-rejection capability is critical to a Brillouin spectrometer. In CBM, this can be achieved by using a two-stage cross-axis virtually imaged phased array (VIPA) configuration, where the light is dispersed twice in orthogonal axes so that noise can be rejected effectively [28, 29]. However, this configuration cannot be used in line-scanning spectrometers as both axes of the VIPA are already in use: one for spatial multiplexing and the other for spectral dispersion. As a result, line-scanning setups are generally running with a single-stage VIPA, whose spectral extinction is too low (e.g., ~30 dB) to suppress non-Brillouin noise. To filter out these excessive noises, current LSBM setups often insert an absorptive gas chamber into the spectrometer as a notch filter. Consequently, the laser source must be frequency-locked to an absorption peak of the atomic gas. To date, this approach has only been demonstrated using a 780-nm tunable laser together with a Rb gas chamber. While effective, it restricts LSBM to near infrared, where the Brillouin signal is much weaker than at shorter wavelength due to the fourth-power scaling of light scattering. In addition, the requirement for tunable laser sources with locking capabilities increases system complexity and hinders widespread adoption of the technology.

Previous work in CBM has demonstrated improved noise rejection by cascading two VIPAs with dispersion axes in parallel, where the first VIPA functions as a spectral filter rather than a spectrum analyzer [30]. Here, we adapted this idea for our inverted co-axial LSBM setup by developing a multiplexed two-stage VIPA spectrometer. Since the parallel VIPAs configuration uses the same optical axis for spectral dispersion, it is well suited for conducting spatial multiplexing on the second axis of VIPAs in line-scanning configuration. In addition, our previous LSBM had a bi-axial optical configuration where the illumination and detection axes were separated, and thus easily suffers from artifacts induced by the refractive index mismatches in heterogeneous samples [26]. To mitigate this effect, we integrated the multiplexed Brillouin spectrometer into a co-axial line-scanning microscope. In this setup, the illumination and detection axes share the same optical path, making the system inherently insensitive to refractive heterogeneity. Furthermore, the inverted configuration eliminates the need of sample mounting and allows biological samples to be directly imaged in culture dishes.

## 2. Experiments

*2.1 Optical setup*

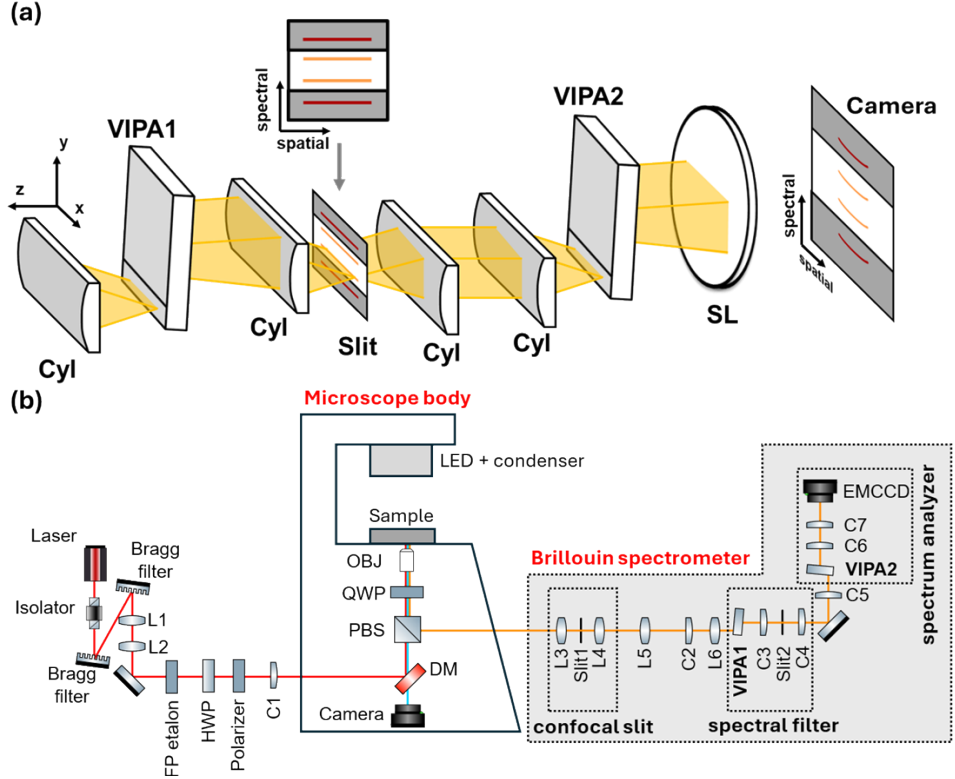

Fig. 1. (a) Principle of the multiplexed two-stage VIPAs spectrometer. Cyl: cylindrical lens; SL: spherical lens. (b) Optical setup of the inverted co-axial LSBM. L1-L5: spherical lenses, C1-C7: cylindrical lenses, FP: Fabry–Pérot, HWP: half-wave plate, DM: dichroic mirror, PBS: polarizing beamsplitter, QWP: quarter-wave plate, OBJ: objective lens; VIPA: Virtually imaged phased array; EMCCD: Electron Multiplying Charge-Coupled Device.

The concept of the multiplexed two-stage VIPA spectrometer is illustrated in Fig. 1(a). The line-scanning Brillouin signal was first coupled into the first VIPA by a cylindrical lens, and the resulting pattern was projected onto a mechanical slit. The slit blocked the laser components while allowing the Brillouin signal to pass through. The transmitted Brillouin signal was then coupled into the second VIPA by a combination of two cylindrical lenses, and the final spectrum was recorded by a camera for analysis. The multiplexing was achieved along the window-length axis of the VIPAs (i.e., x-axis). Fig. 1(b) shows the optical setup of the inverted co-axial LSBM. A 780-nm laser (Toptica, TA Pro) served as the light source. An optical isolator was placed after the laser head to protect it from back-reflected light. The spectrum of the laser was then cleaned using two Bragg filters. A pair of spherical lenses (L1: f = 80 mm, L2: f = 200 mm) expanded the beam to approximately 8.9 mm so that it overfilled the back aperture of the objective lens. A Fabry–Pérot (FP) etalon was used to suppress the side modes of the laser. A half-wave plate (HWP) and a polarizer were used to adjust the beam power and set the polarization to p-polarization. After passing through a cylindrical lens (C1: f = 300 mm), the beam was coupled into an inverted microscope, where it was reflected by a dichroic mirror (DM), transmitted through a polarizing beam splitter (PBS), and focused by the objective lens (OBJ, 40×/0.6) into a line at the focal plane. A quarter-wave plate (QWP) was used to convert the incident illumination beam to circular polarization and the backward-scattered light to s-polarization. A 3D translation stage was used to translate the sample during imaging. An LED light source with a condenser was used for transmission-mode brightfield imaging.

The backward scattered Brillouin signal was reflected by the PBS and guided into the spectrometer. The light first passed through a confocal unit composed of two spherical lenses (L3: f = 200 mm, L4: f = 200 mm) and an adjustable mechanical slit (Slit1). Subsequently, a pair of spherical lenses (L5: f = 150 mm, L6: f = 300 mm) together with a cylindrical lens (C2: f = 50 mm) were used to couple the light into the first VIPA etalon (VIPA1, Free Spectral Range (FSR) = 10 GHz). The dispersion pattern from VIPA1 was projected onto a second slit (Slit2) by a cylindrical lens (C3: f = 100 mm), where the spatially separated laser components were blocked while the Brillouin signal was transmitted (also see Fig. 1(a)). The transmitted light was then reshaped by a cylindrical lens (C4: f = 150 mm), such that VIPA1, C3, Slit2, and C4 together formed a spectral filter. After the filtering unit, the Brillouin signal was coupled into the second VIPA etalon (VIPA2, FSR = 20 GHz) by a cylindrical lens (C5: f = 100 mm). The final dispersion pattern from VIPA2 was projected onto an EMCCD by two cylindrical lenses (C6: f = 200 mm, C7: f = 150 mm), with VIPA2, C6, C7, and the EMCCD together forming the spectrum analyzer.

*2.2 Sample preparation, data acquisition and processing*

The Brillouin shift of the sample is given by:

$$\nu_B = \frac{2n}{\lambda}\sqrt{\frac{M'}{\rho}} \sin\left(\frac{\theta}{2}\right), \qquad (Eq.\ 1)$$

where $\lambda$ is the laser wavelength, $n$ is the refractive index, $\rho$ is the sample's density, $M'$ is the elastic longitudinal modulus, and $\theta$ is the scattering angle. In backward scattering geometry, $\theta = 180°$. Experimentally, the Brillouin shift was determined from the separation between the Stokes and anti-Stokes peaks recorded by the EMCCD camera.

To evaluate the noise suppression capability of the spectrometer, part of the laser light was coupled into the spectrometer through the reflection at the water–glass interface of a Petri dish. Removing C5 and VIPA2 from the optical path allowed the spectral pattern on Mask2 to be directly recorded by the EMCCD camera. one hundred frames of the image were averaged to improve the signal-to-noise ratio (SNR). The recorded image represented the two-dimensional distribution of Brillouin intensity along the spectral and spatial axes (i.e., the x- and y-axes in Fig. 1(a), respectively). The Brillouin spectrum was extracted by summing the intensity over the middle three pixels along the spatial axis, thereby collapsing the spatial dimension and yielding a one-dimensional intensity profile along the spectral axis for analysis.

Deionized (DI) water was used as a reference sample to evaluate the spectral precision of the system. The total illumination power was set to 150 mW, with a camera exposure time of 200 ms. One hundred frames of the Brillouin spectra were recorded. The averaged Brillouin spectrum was obtained by summing the pixel intensity along the spatial axis over the middle three rows. The pixel separation between the Stokes and anti-Stokes peaks was then calculated from the extracted spectrum using Lorentzian fitting. The pixel-to-frequency ratio (PR) was determined as the ratio of the pixel separation to the known Brillouin shift of water. The spectral precision of the spectrometer, in units of GHz, was calculated as the standard deviation of the Stokes–anti-Stokes peak distance over the 100 frames multiplied by the PR. To perform shot-noise analysis, we measured the SNR of the Brillouin signal of water over a range of camera exposure times from 25 ms to 200 ms with the constant input laser power.

The spatial resolution and field-of-view (FOV) of the system was characterized by imaging a 3D-bioprinted cuboid mold (25 μm × 25 μm × 200 μm) fabricated from IP-S photoresist (Nanoscribe) and immersed in DI water of a Petri dish. In-plane and cross-sectional scans were conducted to measure the intensity profile of water's Brillouin signal surrounding the mold. The step size of the stage was 0.25 μm laterally and 0.5 μm axially. The FOV was quantified by the distribution of water's Brillouin signal along the illumination axis. The lateral and axial spatial resolutions was quantified from the intensity profiles across the water–mold interface and the water–glass interface, respectively. In addition, the axial point spread function (PSF) was determined from the Rayleigh scattering intensity of 0.5 μm polystyrene beads moving across the focal plane. The beads were embedded in 2% agarose. For the representative image shown in Fig.7, we prepared a 3D-bioprinted helmet logo mold with same fabrication protocol described earlier. The Brillouin image was acquired by scanning the illumination line with a step size of 1 μm.

## 3. Results

Fig.2(a) and Fig.2(b) show the Brillouin spectrum of water captured by the EMCCD camera when the slit (Slit2) was open and partially closed, respectively. After closing the slit, the laser noise was significantly rejected. To quantify the noise suppression capability, we removed C5 and VIPA2 from the optical path so that the Brillouin pattern on Mask2 can be directly recorded by the EMCCD camera. We then chose the central three pixels along the spatial axis of the Brillouin spectra for analysis. The resulting spectrum of VIPA1 is shown in Fig. 3(a). Specifically, when Slit2 is partially closed to a given extent, the filter provides a corresponding passband with the center passing frequency at half FSR of VIPA1 (here is 5 GHz), and the noise suppression capability was quantified by the transmitted laser power against total input power:

$$\text{Noise suppresion (dB)} = -10\log_{10}\left(\frac{P_\text{residual}}{P_\text{total}}\right), \quad \text{(Eq. 2)}$$

Here, $P_\text{residual}$ is the transmitted laser power, estimated from the spectral integration where the laser leakage is present, and $P_\text{total}$ is the total laser power, estimated from the full integration of the laser spectrum. The resulting laser suppression as a function of passband is shown in Fig. 3(b). When Slit2 was manually set to provide a passband of 2 GHz (passing frequencies: 4 – 6 GHz), the spectral filter achieved a noise suppression of 27 dB, in agreement with the calculated result. Together with the 30-dB noise suppression of VIPA2, we achieved 57-dB noise suppression. Since the Brillouin shift of water is approximately 4 GHz at 780 nm and that of most biological samples generally lies between 4 and 5 GHz, this passband preserves the Brillouin signals of biological samples while effectively suppressing the laser background. The characterization of spectral precision was shown in Fig. 4(a). Using the first pair of Stokes and anti-Stokes peaks (S and AS in Fig. 4(a)), the precision of the water signal was quantified to be 8.3 MHz. The SNR of the Brillouin signal was calculated as a function of input energy on a log-log scale, as presented in Fig. 4(b). The slope of the fitted line was 0.514, indicating the spectrometer runs in shot-noise-limited regime.

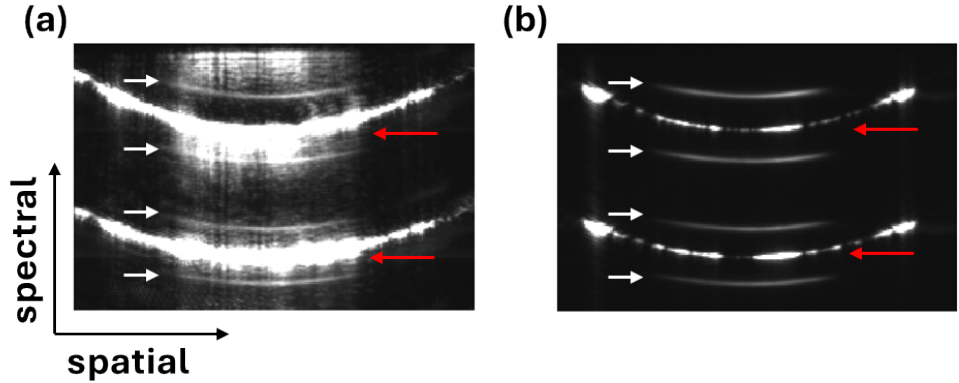

Fig. 2. Brillouin spectra of water when Slit2 is open (a) and partially closed (b). Red arrows indicate the laser noise components, and white arrows indicate the Brillouin signal of water.

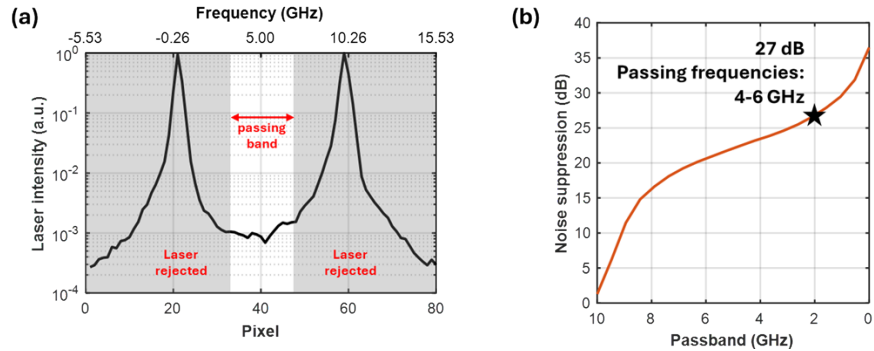

Fig. 3 (a) Laser spectrum of VIPA1. (b) Relationship between theoretical noise suppression capability and the closure of Slit2 (passband). The star marked the experimental result with 2 GHz passband (i.e., 4-6 GHz).

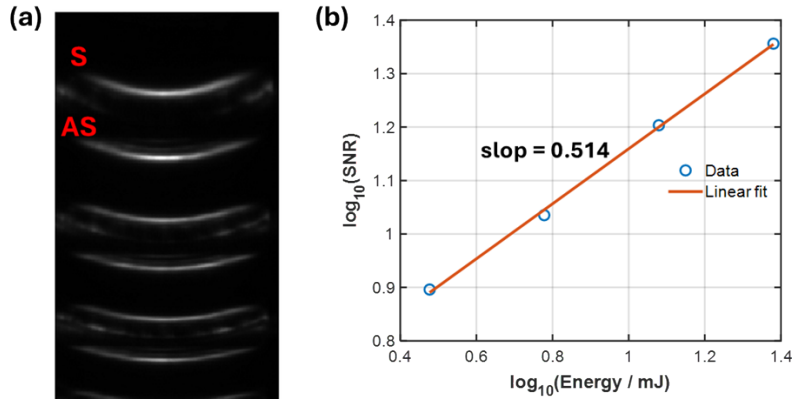

Fig. 4. (a) Brillouin spectrum of water. S: Stokes peak; AS: anti-Stokes peak. The Brillouin shift of water can be calculated from the distance between S and AS. (b) Relationship between input laser energy and the signal's SNR.

The spatial resolution of the system was characterized using a 3D-printed cuboid mold. Fig. 5(a) presents the bright-field image and the corresponding Brillouin intensity image. Since the Brillouin signal of the mold was blocked by the spectral filter, only water's signal was recorded. The FOV of the system was characterized by the distribution of the water-signal intensity along the illumination line, as shown in Fig. 5(b). The results indicate that the system has a $1/e^2$ FOV of 66.5 μm. The intensity profile extracted along the red dashed lines in the Brillouin intensity image was fitted with an error function (erf), and the results are shown in Fig. 5(c) and Fig. 5(d). The full width at half maximum (FWHM) of the derivative suggests the lateral resolution was 1.77 μm and 1.67 μm on the X and Y axis, respectively.

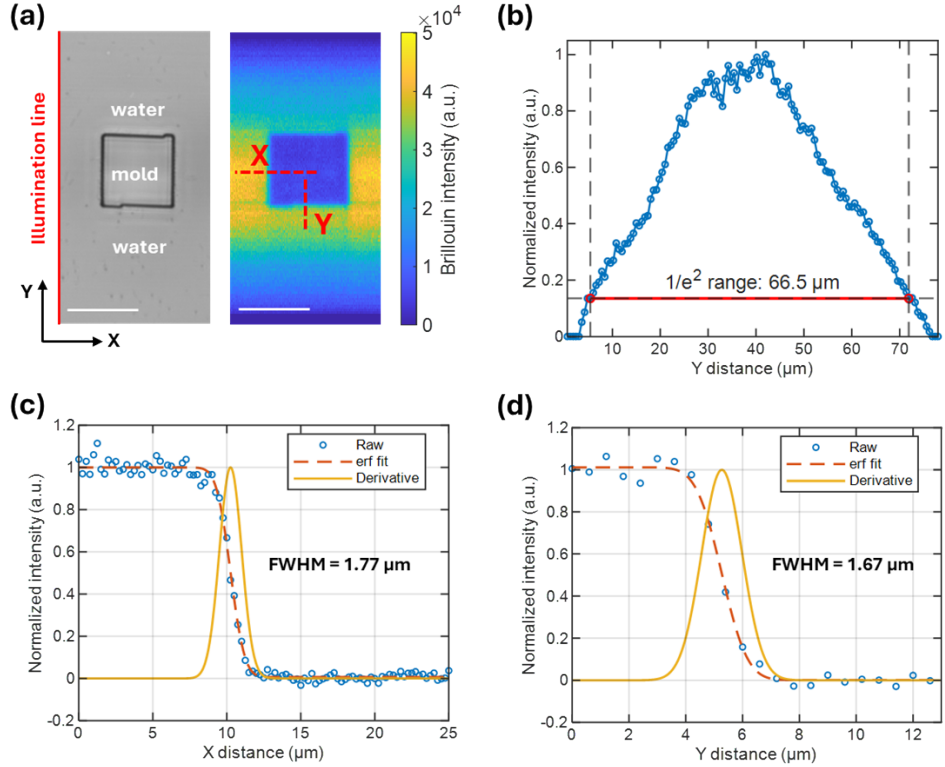

Fig. 5. (a) Bright-field image and Brillouin intensity image of the cuboid mold immersed in water. The solid red line denotes the illumination line, and the red dashed line marks the position used to quantify the lateral resolution. Scale bar: 20 μm (b) Brillouin intensity map of water along illumination line with $1/e^2$ threshold marked. (c) Lateral resolution quantified along the red dashed line X in (a). (d) Lateral resolution quantified along the red dashed line Y in (a).

To quantify the axial resolution, we performed a cross-sectional scan of the mold starting from the bottom of the Petri dish and analyzed the Brillouin signal across the glass-water interface (Fig. 6(a), red dashed line). Unlike the lateral intensity profile, we found that the axial intensity distribution does not follow an ideal erf profile. This is primarily because the line illumination creates a light-sheet-like volume, while the epi-detection geometry allows the confocal slit to collect substantial Brillouin signals from the out-of-focus region. Therefore, instead of applying an erf fit, the raw data were first smoothed, and the axial resolution of 12.91 μm was determined by the FWHM of its derivative, as shown in Fig. 6(b). In contrast, the axial point spread function (PSF) derived from the Rayleigh scattering intensity of a 0.5 μm bead was 3.57 μm, which is about 3.6 times smaller (Fig. 6(c)). We believe that this discrepancy mainly arises from differences in the scattering volume of two methods, which will be discussed in detail later. Finally, we acquired the Brillouin image of a bio-printed mold featuring a helmet logo with a size of 214 μm by 248 μm, as shown in Fig.7. Again, because the Brillouin signal of the mold material (IP-S) was outside the passing band of the VIPA filter and completely blocked, only water signal was recorded. Based on the total acquisition time, the equivalent acquisition speed was determined as 2 ms per pixel.

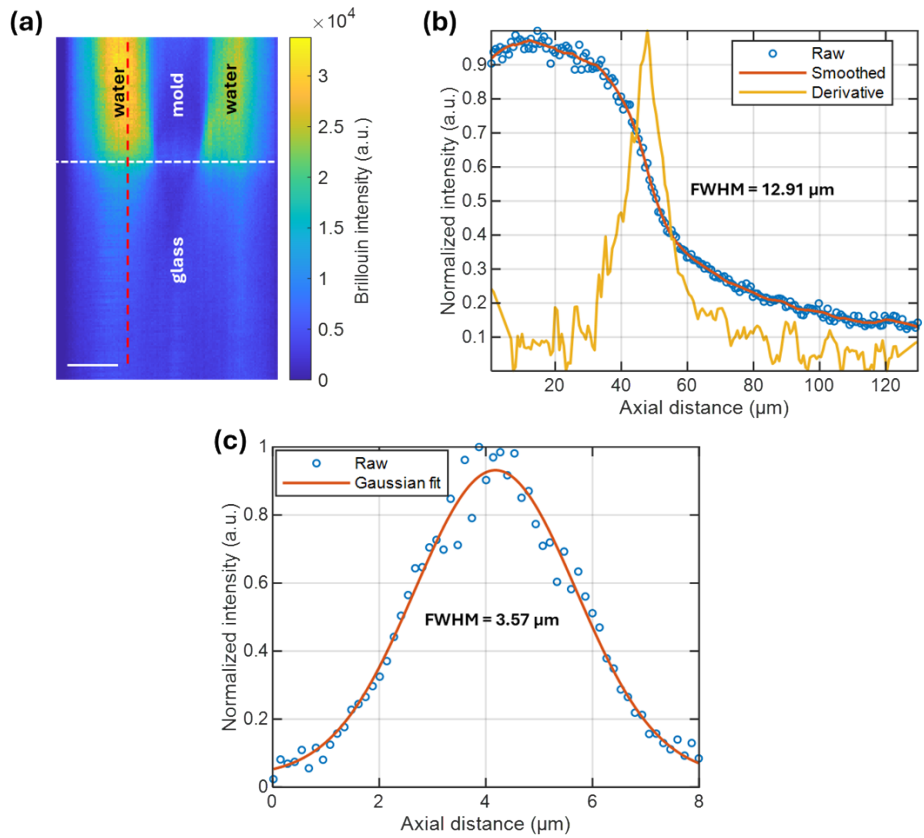

Fig. 6. (a) Brillouin intensity image of water surrounding the cuboid mold. White dash line: glass-water interface; red dash line: the position used to quantify the lateral resolution. (b) Water Brillouin intensity along red dash line in (a). (c) Rayleigh scattering intensity of a 0.5-μm bead moving across the focus.

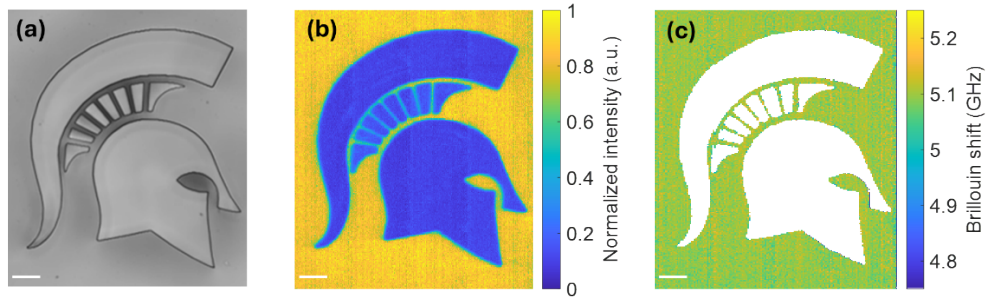

Fig. 7. Representative Bright-field (a), normalized water Brillouin signal intensity (b), and Brillouin shift image (c) obtained from co-axial line-scanning Brillouin microscope. Scale bar: 25 μm.

## 4. Discussion & Conclusion

Regarding the spectral filter unit in Fig.1(b), the selection of VIPA1's FSR is critical. Since the noise-rejection capability relies on how much Slit2 can be closed without blocking the Brillouin signal, a general practice is to align the sample's Brillouin shift as closely as possible with the

VIPA's central transmission frequency. For a 780-nm laser source and backward scattering geometry, the Brillouin shift of water is around 5 GHz. Since the Brillouin shifts of many biological specimens typically fall within the 5–6 GHz range, we selected a FSR of 10 GHz for VIPA1. In this case, the Stokes and anti-Stokes peaks of water largely overlap, while the two peaks of biological specimens remain very close to each other. As a result, Slit2 can be closed tightly, leading to enhanced rejection of non-Brillouin noise. In contrast, when the FSR of VIPA1 was 15 GHz, the central frequency became 7.5 GHz. To ensure the transmission of Brillouin signal with shifts of 5–6 GHz, a minimum passband of 5 GHz was required, corresponding to passing frequencies ranging from 5 to 10 GHz. Consequently, only 18 dB of noise could be suppressed (Fig. S1). This selection principle can also be applied to VIPAs working at different wavelengths. For instance, with a 660-nm laser, biological samples exhibit Brillouin shifts of 6–7 GHz. Therefore, an FSR of 13.5 GHz will be an optimal choice.

Currently, the spectral precision was quantified by using the first set of Brillouin peaks (i.e., S and AS in Fig. 4). It is worth noting that this can be further improved by including Brillouin peaks from multiple orders for analysis. In principle, by averaging the results from multiple orders, the final precision, $\sigma_{\text{average}}$, is given by:

$$\sigma_{average} = \frac{1}{n}\sqrt{\sum_{i=1}^{n} \sigma_i^2}, \tag{Eq. 3}$$

where $\sigma_i$ is the precision of the $i^{\text{th}}$ order. Table S1 summarizes the precision calculated from the first to the third order, together with the precision obtained by averaging all three orders. It is shown that the averaged precision of three orders was 6.1 MHz, which is close to the calculated value of 5.9 MHz using Eq. 3.

As shown in Fig.6, we observed a remarkable difference between the axial resolution characterized by the bead-based optical PSF and that quantified from the intensity profile of Brillouin signals (3.57 μm vs. 12.91 μm). This difference primarily arises from the distinct scattering volume in the two measurements and is further amplified by the confocal slit used in line-scanning setups. In bead-based PSF measurement, the collected signal is from a point source and highly localized. In contrast, the intensity profile of Brillouin signals is a sum of the scattering sources from everywhere within an extended volume, which is similar to the scenario of densely labelled fluorescent sample (i.e., fluorescence sea) (Fig. S2) [31]. Consequently, Brillouin signals from locations in out-of-focus plane but off-axis will be collected by the spectrometer and thus degrades the axial resolution. In addition, compared to a confocal pinhole, the confocal slit only rejects out-of-plane light along one dimension (i.e., the slit-width axis). The freely passing light along the other dimension (i.e., the slit-length axis) causes severe crosstalk between adjacent pixels within the scattering volume, which further degrades the axial resolution quantified from the Brillouin signals. To demonstrate the impact of the confocal slit, we modified the illumination from a line to a focused spot by removing the cylindrical lens C1 of Fig. 1. The recorded Brillouin spectrum of water exhibits two long tails, which are from out-of-plane light (Fig. S3a). Notably, the contribution from the out-of-plane light could be quantified based on their spatial location on the camera. For instance, excluding the tails in the postprocessing yielded an axial resolution of 7.5 μm, whereas including them degraded the resolution to 10.2 μm (Fig. S3b and Fig. S3c). Alternatively, these tails disappeared when the confocal slit was replaced with a pinhole (Fig. S3d), and the axial resolution remained 8.1 μm regardless of whether the corresponding tail regions were included or not (Fig. S3e and Fig. S3f). This confirms the tails originated from the out-of-plane light that bypasses the confocal slit. In LSBM setups, the illumination line can be approximated as a continuous array of closely spaced focal spots. The tails generated at each spot accumulate as background noise over the adjacent in-plane signal. As a result, this significantly reduces the contrast and degrades the overall axial resolution. Taken together, our results suggest that characterizing the Brillouin intensity profile across the interface of two materials provides a more accurate quantification of the LSBM's axial resolution than bead-based PSF measurements.

Despite its limited sectioning ability, our LSBM setup maintains the advantages of rapid acquisition speed and strong laser-noise suppression provided by the multiplexed two-stage VIPA spectrometer. Importantly, the noise suppression capability can be further enhanced by cascading a second filter unit in the spectrometer if desired. These features make it well suited for probing the viscoelastic properties of inhomogeneous, highly scattering, large-size samples such as embryonic tissues [32-34].

In conclusion, we have developed a coaxial line-scanning Brillouin microscope integrated with a multiplexed two-stage VIPA spectrometer. The spectrometer can suppress non-Brillouin noise by 57 dB, allowing mechanical imaging of bio-printed phantoms. Since a gas chamber is not needed for noise suppression, this approach removes current constraints on laser source selection and makes LSBM readily adaptable to other wavelengths (e.g., 532 nm and 660 nm), where commercial lasers are widely available and spontaneous Brillouin scattering is more efficient.

**Acknowledgements**
The authors thank Y. Liu and M. Li at Michigan State University for helping prepare the 3D-printed molds for Brillouin imaging. This project is funded by the National Institutes of Health (R21HD112663) and the National Science Foundation (CBET- 2546314).

**Disclosures**
A provisional patent application related to this research has been filed. The authors declare no other conflicts of interest.

**Data availability**
Data underlying the results presented in this paper can be obtained from the authors upon reasonable request.

**Supplemental document**
See Supplement 1 for supporting content.